\documentstyle[prl,aps,multicol]{revtex}
\input{epsf}
\begin{document}
\draft

\title{Lasing Threshold and Mode Competition in Chaotic Cavities}

\author{T.~Sh.~Misirpashaev$^{a,b}$ and C.~W.~J.~Beenakker$^a$}
\address{$^a$Instituut-Lorentz, University of Leiden,
                 P.O. Box 9506, 2300 RA Leiden, The Netherlands\\
$^b$Landau Institute for Theoretical Physics, 2 Kosygin Street,
Moscow 117334, Russia}
\date{August 22, 1997}
\maketitle

\begin{abstract}

The lasing threshold is studied of a multi-mode chaotic cavity 
(linear size $D$ $\gg$ wavelength $\lambda$)
coupled to the outside through a small hole 
(linear size $d \ll \lambda$).
For sufficiently weak absorption by the boundaries,
the statistical distribution 
of the threshold is wide, its mean value
being much less than the pumping rate needed to compensate 
the average loss. 
The average number $\langle N_{nc}\rangle\gg1$ of non-competing
excited modes is proportional to the square root of the pumping rate.
We use the classical model of spatial hole burning to account for  
mode competition and find a reduction in the average number of excited modes to 
$\langle N\rangle=3^{1/3} \langle N_{nc}\rangle^{2/3}$.
  
\end{abstract}

\begin{multicols}{2}
\narrowtext
\section{INTRODUCTION}
Incorporation of quantum optical effects is a necessary 
and interesting 
extension of active ongoing research on 
multiple scattering of electromagnetic waves 
in random media \cite{Lag96}. It becomes particularly important
when the medium is active, as is the case 
in experimentally realized ``random lasers''
\cite{Law94,Sha94,Wie95}. Quantum effects
have been largely ignored in many publications devoted
to propagation in disordered amplifying waveguides 
\cite{Pra94,Zha95,Bee96,Paa96},
in which
only amplified stimulated emission of external incoming 
flux but not of spontaneously emitted internal noise 
was taken into account (Ref.~\onlinecite{Zyu95} being 
a notable exception). Amplified internal noise can lead to
excitation of low-threshold lasing modes of the waveguide, 
making practical use of amplifying waveguides problematic.
The difficulty of the waveguide geometry is the onset of localization.
In this paper we consider a simpler cavity geometry, 
which does not show localization, but retains 
two essential features of the problem: large sample-to-sample
fluctuations and spontaneous emission. 

A complete description 
of the fluctuations is possible in the universal regime
characterized by a chaotic pattern of classical trajectories.
We assume that the cavity (volume $V\simeq D^3$) is confined 
by conducting walls, filled with a lasing medium
(central frequency of the gain profile $\omega_0$),
and coupled to external detectors via one or several small 
holes.
 It was demonstrated recently that a non-integrable 
shape of the resonator can 
significantly affect its lasing properties \cite{Mek95}.  
Chaoticity of classical trajectories can be achieved either 
by a peculiar shape of the resonator \cite{Bun96,Alt95,Alt97}, or
by a small amount
of disorder scattering. We will speak about ``chaotic cavities,''
meaning any of the two mechanisms responsible for the onset of chaos.  

We restrict ourselves to 
the case of well-resolved cavity modes, which means that 
1) resistive loss $\gamma_*$ in the cavity walls is less than the mean modal
spacing $\delta\omega_0=\pi^2 c^3/\omega_0^2 V$, 2) characteristic size of the holes $d$ is smaller than the wavelength $\lambda_0=2\pi c/\omega_0$. 
Mean loss $\gamma_0$ through a small hole
was calculated by Bethe \cite{Bet44},
\begin{equation}
\gamma_0\simeq\frac{cd^6}{\lambda_0^4V}, \qquad d\ll\lambda_0,
\label{Bethe}\end{equation}
so that $\gamma_0/\delta\omega_0\simeq(d/\lambda_0)^6\ll1$.
Note that the loss (\ref{Bethe}) is not proportional to the area of
the hole. It is in fact much smaller than one might guess by 
extrapolating the dependence $\gamma_0\simeq cd^2/V$ valid
for $d\gg\lambda_0$.
The effect of sample-to-sample fluctuations
is pronounced only if $\gamma_*\ll\gamma_0$. 
This regime is experimentally accessible, as
was demonstrated by a recent 
series of experiments on microwave cavities with superconducting Niobium
walls \cite{Alt95,Alt97}.  
  
Each act of spontaneous emission in a pumped cavity 
is a source of radiation into some cavity mode. 
Classical condition of the lasing threshold in 
a given cavity mode is satisfied if the gain due to stimulated
emission equals the loss. 
Threshold for the cavity is the smallest value of the pumping
rate at which threshold is attained for one of the modes.
The questions we ask are: What is the threshold rate of
pumping? How many lasing modes can coexist for a given  
pumping rate above the threshold?    
The problem of spectral content
of outgoing radiation has been widely studied for integrable 
cavities of definite shape.
Considering arrays of chaotic cavities of slightly varying shape
or with different configurations of scatterers
we address the problem 
statistically and compute the probability of lasing,
the distribution of the threshold, and the average number 
of excited modes.

Trivially, gain greater than mean loss $\gamma_0$ will be
on the average sufficient to ensure lasing, while
gain smaller than $\gamma_*$ will never suffice. 
The mean loss from a tiny hole is small. We argue that 
the actual average threshold can still be many orders of
magnitude smaller.  Each individual cavity
exhibits a well-defined threshold but its statistical distribution
is wide. In Section~\ref{gamma*=0} we compute this distribution for 
the idealized case $\gamma_*=0$. Effects of non-zero resistivity
of the walls are discussed in Section~\ref{gamma*>0}.
Section~\ref{numberofmodes} is devoted to the computation of the average
number of excited modes above the threshold. We conclude in Section~\ref{conclusions}.

\section{DISTRIBUTION OF LASING THRESHOLD} 
\label{gamma*=0}
We assume that the line of spontaneous emission is homogeneously 
broadened and has Lorentzian shape
with central frequency $\omega_0$
and  width $2\Omega$.
Let $\psi_i(\vec r)$ be the amplitude of a mode
of the closed cavity 
at frequency $\omega_i$, normalized according to
$\int\! d\vec r\,\psi_i^2(\vec r)=V$.
(For simplicity we neglect polarization dependent phenomena and work with
real scalar field amplitudes.)
In the presence of weak coupling to the outside world the modes acquire finite widths, 
$\gamma_i$. We assume two sets of conditions:

\begin{equation}
\gamma_0 \ll \delta\omega_0 < \Omega \ll \omega_0,
\label{ineq1}\qquad
d \ll \lambda_0 \ll D.
\label{ineq2}\end{equation}
An especially important role is played by the inequality 
$\gamma_0 \ll \delta\omega_0$, which is implied by $d \ll \lambda_0$.
It ensures that the modes of the open cavity are well-defined and do
not differ significantly from those of the closed one.
In this Section we consider the idealized case in which there is no loss
in the walls of the cavity ($\gamma_*=0$). 

In a chaotic cavity the modes $\psi_i(\vec r)$ can be modeled 
as random superpositions of plane waves
\cite{Ber77}. 
(Validity of this model has been checked experimentally
in microwave cavities \cite{Alt95,Pri95}.)
This implies a Gaussian 
distribution for $\psi_i(\vec r)$ at any point $\vec r$. 
The corresponding distribution for $\psi_i^2(\vec r)$ is 
called the Porter-Thomas distribution \cite{Bro81}.
Loss from 
a small hole located at $\vec r$ is proportional to 
$\left[\nabla_{\vec n}\psi_i(\vec r)\right]^2$ 
(with $\nabla_{\vec n}\psi_i(\vec r)$ the derivative in the
direction normal to the surface of the hole) and has the same 
Porter-Thomas distribution, which was directly probed in 
the experiments of Ref.~\onlinecite{Alt95}. 
More generally, the distribution of
normalized modal widths $y=\gamma_i/\gamma_0$ 
in a cavity with $\nu$ holes is given by 
the $\chi^2$-distribution with $\nu$ degrees of freedom (normalized to $1$),

\begin{equation}
P_\nu(y)=\frac{(\nu/2)^{\nu/2}}{\Gamma(\nu/2)}y^{-1+\nu/2}\exp(-\nu y/2).
\label{chi2}\end{equation}
We assumed that loss from different holes is independent, which is true
 provided their separation is larger than $\lambda_0$.
For small integer $\nu$, the distribution  (\ref{chi2}) is wide. 
The single-hole case $\nu=1$ looks especially promising from the point of view of 
low-threshold lasing because $P_1(y)=\exp(-y/2)/\sqrt{2\pi y}$
grows with decreasing $y$.

To grasp the picture we first confine ourselves to
a subset of cavity modes located
near $\omega_0$. We neglect fluctuations of their frequencies
and assume that the modes are equidistant, 
$\omega_m=\omega_0+\delta\omega_0 m$, $m=0,\pm1,\pm2,\ldots$  
We denote by $R_{p_0}$ a reference pumping 
rate necessary to provide
gain equal to the mean loss $\gamma_0$  at frequency $\omega_0$, and 
introduce the reduced pumping rate 
${\varepsilon}=R_p/R_{p_0}$, assumed ${}\ll1$. 
Loss of different modes is uncorrelated and
distributed according to Eq.~(\ref{chi2}) while gain diminishes with increasing 
difference $|\omega-\omega_0|$ according to the Lorentzian
\begin{equation}
g_0(\omega)=\gamma_0{\varepsilon}\left[1+
(\omega-\omega_0)^2/\Omega^2\right]^{-1}.
\label{modgain}\end{equation}
It follows that
the probability $p_\nu({\varepsilon})$ of there being no lasing
mode at the pumping rate ${\varepsilon}$ is given by
\begin{equation}
p_\nu({\varepsilon})=\prod_m
\left(1-\int_0^{g_0(\omega_m)/\gamma_0}\!\!\!dy\,P_\nu(y)\right).
\label{p1}\end{equation}
For ${\varepsilon}\ll 1$, the upper limit of the integral is also 
$\ll 1$, and we can replace $P_\nu(y)$ by its leading behavior 
at small $y$, $P_\nu(y)\propto y^{-1+\nu/2}$, which yields

\begin{eqnarray}
&&p_\nu({\varepsilon})\approx\prod_m\Bigl(1-
\frac{C_\nu{\varepsilon}^{\nu/2}}{[1+m^2(\delta\omega_0/\Omega)^2]^{\nu/2}}\Bigr)\cr
&&\hphantom{xxx}\approx\exp\left(-C_\nu{\varepsilon}^{\nu/2}
\sum_m\left[1+m^2(\delta\omega_0/\Omega)^2\right]^{-\nu/2}\right),\label{p2}\\
&&\hphantom{xxx}C_\nu=(\nu/2)^{-1+\nu/2}[\Gamma(\nu/2)]^{-1}.\label{Cnu}
\end{eqnarray}

Because the summand decays as $m^{-\nu}$  we find that for $\nu>1$
the leading behavior of 
the probability
of no-lasing 
is determined by the modes with $|m|\lesssim\Omega/\delta\omega_0$,
\begin{equation}
p_\nu({\varepsilon})\approx\exp\left[-\tilde C_\nu(\Omega/\delta\omega_0)
{\varepsilon}^{\nu/2}
\right].
\label{1minp}\end{equation}
Here $\tilde C_\nu=\sqrt{\pi}C_\nu\Gamma((\nu-1)/2)/\Gamma(\nu/2)$
for $\nu>1$ (below we will separately define $\tilde C_1$). 
Modes far from $\omega_0$ have 
negligible chance to get excited and need not be taken into account.
On the contrary, for $\nu=1$  all cavity modes,
including those very far from $\omega_0$, contribute to the probability. 

To treat the contribution of distant modes for $\nu=1$ correctly, we must 
account for several factors
which we could ignore for $\nu>1$.
 1)~The spectral density cannot be replaced
by its value $\rho_0=1/\delta\omega_0$ at $\omega=\omega_0$. Instead
$\rho(\omega)=\rho_0\omega^2/\omega_0^2$. 2)~The mean loss
is frequency dependent, $\bar\gamma(\omega)=\gamma_0\omega^4/\omega_0^4$, 
cf.\ Eq.~(\ref{Bethe}). 3)~The Lorentzian~(\ref{modgain})
for the amplification rate is an approximation valid only in the vicinity 
of $\omega_0$. A correct expression for the gain $g(\omega)$ must be even in 
$\omega$ to comply with the symmetry 
$\chi(\omega)=\chi^*(-\omega)$
of the dielectric susceptibility $\chi$.
It includes contributions of both poles $\pm\omega_0+i\Omega$ 
and reads 
\begin{equation}
g(\omega)=\frac{4\omega^2\gamma_0{\varepsilon}\Omega^2}
{(\omega^2-\omega_0^2)^2+2(\omega^2+\omega_0^2)\Omega^2+\Omega^4}.
\label{corgain}\end{equation}

\begin{figure}
\hspace*{18truemm}
\epsfxsize=0.6\hsize
\epsfysize=0.12\vsize
\epsffile{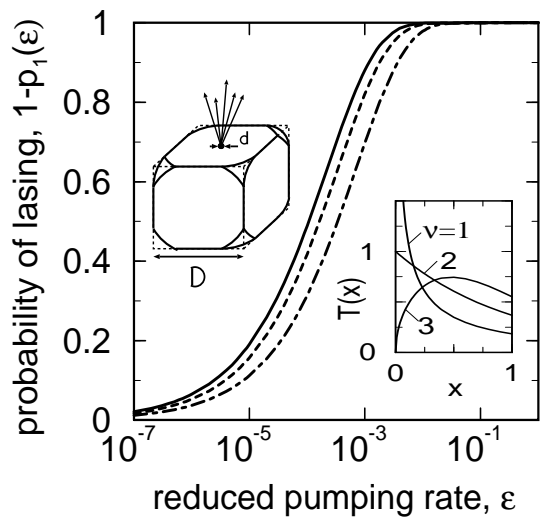}
\vglue 15truemm
\caption{Probability of lasing $1-p_1({\varepsilon})$ versus reduced pumping rate 
${\varepsilon}$ with $p_1({\varepsilon})$ 
given by Eq.~(\protect\ref{pexa}) ($\omega_0/\Omega=10$, 
$\Omega/\delta\omega_0=10$). Thick lines
are for different ratios $D/d$ corresponding to 
different numbers $M\simeq(D/d)^3$ 
of relevant cavity modes (dot-dashed line $M=10^2$,
dashed line $M=10^3$, solid line $M=10^4$).
 Left inset shows an example of chaotic cavity. 
Chaotic behavior of classical trajectories in 
this particular ``die'' shaped cavity was shown
in Ref.~\protect\cite{Bun96}.
Radiation is confined inside
by means of ideally conducting walls and can leave the cavity only
through a tiny hole. Right inset shows the probability 
distribution of the lasing threshold 
$T_\nu(x)=(\nu/2)x^{-1+\nu/2}\exp(-x^{\nu/2})$,
with $x$ related to ${\varepsilon}$ by
$x={\varepsilon}(\tilde C_\nu\Omega/\delta\omega_0)^{2/\nu}$,
for different number of holes
$\nu=1,2,3$. }
\label{f_cavity}\end{figure}

Taking these three factors into account and replacing
the discrete sum by an integral, the probability
of no-lasing is given by
\begin{equation}
p_\nu({\varepsilon})=\exp\left(\int_0^{\omega_{max}}\!\!\!\!
d\omega\,\rho(\omega)\ln\int_{g(\omega)/\bar\gamma(\omega)}^\infty
\!\!\!\!dy\,P_\nu(y)\right).
\label{pgeneral}\end{equation}
For $\nu>1$ this leads to Eq.~(\ref{1minp}), the ultraviolet cutoff
$\omega_{max}$ being irrelevant. For $\nu=1$ we get
\begin{equation}
p_1({\varepsilon})=\exp\left(\int_0^{\omega_{max}}\!\!\!\!
d\omega\,\rho(\omega)\ln\left[1-\hbox{\rm erf}
\sqrt{\frac{g(\omega)}{2\bar\gamma(\omega)}}\right]\right), 
\label{pexa}\end{equation} 
where $\hbox{\rm erf}(z)=(2/\sqrt{\pi})\int_0^z dx\exp(-x^2)$
is the error function.
The main logarithmic contribution of type $\int\! d\omega/\omega$ 
to the integral in Eq.~(\ref{pexa})
 comes from large values of $\omega$.
The ultraviolet cutoff $\omega_{max}\simeq 2\pi c/d$ appears because loss 
of high frequency modes with 
$\lambda<d$ no longer exhibits the strong fluctuations of Eq.~(\ref{chi2}).
Beyond the cutoff classical ray optics applies, leading
to a narrowly peaked distribution of the loss around the value $cd^2/V\gg\gamma_0$.
Because we are considering the case ${\varepsilon}\ll 1$ 
in which the gain is much smaller
than the average loss $\gamma_0$, the high frequency modes cannot be excited.
It follows that the only relevant cavity modes
are those with frequencies smaller than $\omega_{max}$. 
Their number
$M\approx\omega_{max}^3/3\omega_0^2\delta\omega_0\simeq(D/d)^3$ is ${}\gg1$.
From Eq.~(\ref{pexa})
the probability of no-lasing $p_1({\varepsilon})$ can be cast in the form of
Eq.~(\ref{1minp}) with the coefficient 
$\tilde C_1=(8/\pi)^{1/2}\ln(\omega_{max}/\Omega)$ weakly dependent 
on the frequency cutoff $\omega_{max}$. 
Fig.~\ref{f_cavity} shows that the probability of lasing $1-p_1({\varepsilon})$ can be 
reasonably large even for extremely small values of
the reduced pumping rate~${\varepsilon}$.

The quantity $1-p_\nu({\varepsilon})$ is the fraction of lasing cavities in 
an array at a given pumping rate ${\varepsilon}$. It is directly related to 
the probability distribution 
$T_\nu({\varepsilon})$ of the lasing threshold.
Obviously 
$\int_0^{\varepsilon} d{\varepsilon}' T_\nu({\varepsilon}')=1-p_\nu({\varepsilon})$, 
hence $T_\nu({\varepsilon})=-dp_\nu({\varepsilon})/d{\varepsilon}$. We find from 
Eq.~(\ref{1minp}) that 
\begin{equation}
T_\nu({\varepsilon})=\case{1}{2}\nu\tilde C_\nu(\Omega/\delta\omega_0)
{\varepsilon}^{-1+\nu/2}
\exp\left[-\tilde C_\nu(\Omega/\delta\omega_0){\varepsilon}^{\nu/2}\right].
\label{thd}\end{equation}
(Deviations which arise at ${\varepsilon}\gtrsim1$ are unimportant.)
The distribution is wide and  
in the single-hole case $\nu=1$ diverges as ${\varepsilon}\to{+}0$ (see
right inset of Fig.~\ref{f_cavity}). 
The average reduced threshold reads
$\langle{\varepsilon}_\nu\rangle=\Gamma(1+2/\nu)(\tilde C_\nu\Omega/\delta\omega_0)^{-2/\nu}.$ It is smallest for 
$\nu=1$ and is indeed much smaller than 1. 

\section{EFFECTS OF NON-ZERO WALL RESISTIVITY}
\label{gamma*>0}  

\begin{figure}
\hspace*{15truemm}
\epsfxsize=0.65\hsize
\epsfysize=0.11\vsize
\epsffile{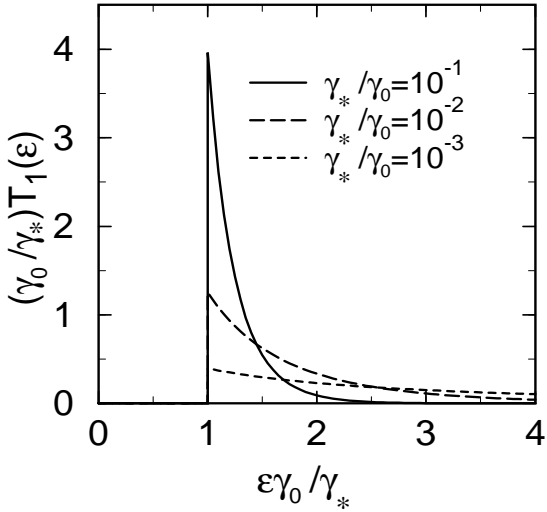}
\vglue 15truemm
\caption{Probability distribution of the lasing threshold in a cavity
with small absorption in the boundary and a single hole ($\nu=1$), 
computed as $-dp_1({\varepsilon})/d{\varepsilon}$ from  
Eq.~(\protect\ref{pviaf}). We chose
$\Omega/\delta\omega_0=10$ and took three values of 
$\gamma_*/\gamma_0$ such that 
$\gamma_*/\gamma_0$ is much smaller, equal, or
much greater than $(\delta\omega_0/\Omega)^{2/\nu}=10^{-2}$.
}
\label{fgamma*}\end{figure}

A non-zero loss $\gamma_*$ from the resistivity of the cavity walls
modifies the functions (\ref{pgeneral}--\ref{thd}) by suppressing
lasing for ${\varepsilon}<\gamma_*/\gamma_0$. The distribution of the lasing
threshold remains wide, as long as $\gamma_*/\gamma_0\ll1$, as we now show.
Instead of Eq.~(\ref{pgeneral}) we have
\begin{equation}
p_\nu({\varepsilon})=\exp\left(\int_{\omega_-}^{\omega_+}\!\!\!\!
d\omega\,\rho(\omega)\ln\int_{(g(\omega)-\gamma_*)/\bar\gamma(\omega)}^\infty
\!\!\!\!dy\,P_\nu(y)\right),
\label{pgeneral*}\end{equation}
where $\omega_-<\omega_+$ are the two positive frequencies such that
$g(\omega_\pm)=\gamma_*$. A non-zero value of $\gamma_*$ 
reduces the relevant frequency
range to a narrow window around $\omega_0$. 
Therefore, the modifications 1)-3)
of the previous Section become unnecessary even for the case $\nu=1$.
Using the simple Lorentzian~(\ref{modgain}) for $g(\omega)$,  
instead of the more complicated expression~(\ref{corgain}),
we find $\omega_\pm=\omega_0\pm
\Omega({\varepsilon}\gamma_0/\gamma_*-1)^{1/2}$. Neglecting the
$\omega$-dependence of $\rho(\omega)$, $\bar\gamma(\omega)$ and using 
the small-argument behavior of the probability function $P_\nu(y)$,
we reduce Eq.~(\ref{pgeneral*}) to 
\begin{equation}
p_\nu({\varepsilon})=
\exp\Bigl[-C_\nu(\Omega/\delta\omega_0)
(\gamma_*/\gamma_0)^{\nu/2} 
f_\nu({\varepsilon}\gamma_0/\gamma_*-1)
\Bigr],
\label{pviaf}\end{equation}
where $C_\nu$ is the numerical coefficient introduced in Eq.~(\ref{Cnu})
and 
\begin{equation}
f_\nu(z)=\sqrt{z}\int_{-1}^1 dy\left(\frac{1-y^2}{y^2+1/z}\right)^{\nu/2}
\label{fnu}\end{equation}
can be expressed in terms of a hypergeometric function.
In Fig.~\ref{fgamma*} we have plotted the distribution 
of the lasing threshold, $T_\nu({\varepsilon})=-dp_\nu({\varepsilon})/d{\varepsilon}$,
for $\nu=1$ and different values of $\gamma_*/\gamma_0$.
We will analyze two limiting regimes.
\vglue 10truemm

In the regime ${\varepsilon}\gamma_0/\gamma_*\gg1$ and for $\nu>1$ 
we recover the expression~(\ref{thd}) with the same constant $\tilde C_\nu$.
The value of $\tilde C_1=C_1\ln({\varepsilon}\gamma_0/\gamma_*)$ is
 different because of the different cutoff
mechanism. Instead of having a weak logarithmic dependence on $\omega_{max}$ it 
exhibits a weak logarithmic dependence on the pumping rate ${\varepsilon}$. This limiting case is statistically dominant if 
$\gamma_*/\gamma_0\ll(\delta\omega_0/\Omega)^{2/\nu}$, 
because then the corrections to Eq.~(\ref{thd})
at ${\varepsilon}\lesssim\gamma_*/\gamma_0$
have negligible statistical weight.  

In the opposite regime, ${\varepsilon}\gamma_0/\gamma_*-1\ll1$, the threshold
distribution differs significantly from Eq.~(\ref{thd}),
\begin{eqnarray}
&&T_\nu({\varepsilon})=
\case{1}{2}(1+\nu)A_\nu(\Omega/\delta\omega_0)(\gamma_0/\gamma_*)^{1/2}
({\varepsilon}-\gamma_*/\gamma_0)^{\nu-1\over2}\nonumber\\
&&\;\mbox{}\times\exp\left[-A_\nu(\Omega/\delta\omega_0)(\gamma_0/\gamma_*)^{1/2}
({\varepsilon}-\gamma_*/\gamma_0)^{1+\nu\over2}\right]
\label{case2}\end{eqnarray}
(with a numerical coefficient
$A_\nu=\sqrt{\pi}C_\nu\Gamma(1+\nu/2)/\Gamma(3/2+\nu/2)$).
This regime is statistically dominant if 
$\gamma_*/\gamma_0\gg(\delta\omega_0/\Omega)^{2/\nu}$. 
The mean value of threshold is now
close to $\gamma_*/\gamma_0$, but there are large fluctuations
towards larger ${\varepsilon}$.

\section{AVERAGE NUMBER OF EXCITED MODES}
\label{numberofmodes}
In this Section we focus on the number of lasing modes 
beyond the lasing threshold
for $\nu=1$ assuming $\gamma_*=0$. We assume that the parameters are such
that many modes are above the threshold. This requires, in particular,
${\varepsilon}\gamma_0/\gamma_*\gg1$. In this case 
a non-zero value of $\gamma_*$ only leads to a redefinition
of $\tilde C_1$ because of the different cutoff mechanism.
If the modes did not compete we could compute the average
number of excited modes $\langle N_{nc}\rangle$ as
\begin{equation}
\langle N_{nc}\rangle=
\int_0^{\omega_{max}}\!\!\!
d\omega\,\rho(\omega)\hbox{\rm erf}
\sqrt{\frac{g(\omega)}{2\bar\gamma(\omega)}}.
\end{equation}
For ${\varepsilon}<1$ it is given by 
$\langle N_{nc}\rangle=
\tilde C_1(\Omega/\delta\omega_0) {\varepsilon}^{1/2}$.
However, the modes do compete 
for a homogeneously broadened line because
one of the modes can deplete the inversion, preventing another mode
from being excited \cite{Sie86}. Multi-mode operation is 
still possible if different excited modes deplete the inversion
in different spatial regions of the cavity \cite{Hak63,Tan63}. 
We assume this mechanism of multi-mode generation, called spatial hole burning \cite{Hak85}.

Let $n_i$, ${\cal N}(\vec r)$ denote the number of photons in the mode $i$ 
and the density of population inversion between the lasing levels.
Semiclassical rate equations read 
\begin{eqnarray}
&&\frac{dn_i}{dt}=-\gamma_in_i+W_in_i\int\! d\vec r\,
\psi_i^2(\vec r){\cal N}(\vec r), \\
&&\frac{d{\cal N}(\vec r)}{dt}={\varepsilon} R_{p_0}/V-w{\cal N}(\vec r)-{\cal N}(\vec r)
\sum_iW_in_i\psi_i^2(\vec r). 
\end{eqnarray}
Here $w$ is the non-radiative
decay rate and $W_i$ is the rate of stimulated emission into 
mode $i$. The constant $W_i$ is related to the gain (\ref{corgain}) 
in the corresponding
mode, $W_i=wg(\omega_i)/{\varepsilon} R_{p_0}$.

We restrict ourselves to a steady state solution. Eliminating the
equilibrium  population inversion density ${\cal N}(\vec r)$, we get the following 
set of equations for the equilibrium mode populations $n_i$,
\begin{equation}
\left(-\gamma_i+{\varepsilon} R_{p_0}
W_i\int\!\! \frac{d\vec r}{V}\frac{\psi_i^2(\vec r)}
{w+\sum_j W_j n_j \psi_j^2(\vec r)}\right)n_i=0.
\label{nonlin}\end{equation}
Vanishing $n_i$'s correspond to non-excited modes. Let there be $N$
excited modes, $i_1,i_2,\ldots i_N$.   
We assume that we are not far beyond threshold, so that  
$w\gg\sum_j W_j n_j 
\psi_j^2(\vec r)$, and we may expand the denominator in Eq.~(\ref{nonlin}).  
We arrive at the following system of linear equations ($k=1,\ldots N$)
\begin{equation}
\frac{1}{{\varepsilon} R_{p_0}}
\sum_{l=1}^N A_{i_ki_l}g(\omega_{i_l})n_{i_l}=1-\frac{\gamma_{i_k}}
{g(\omega_{i_k})},
\label{lin}\end{equation}
subject to a constraint $n_{i_k}>0$.
Coefficients $A_{i_ki_l}$ are given by
\begin{equation}
A_{i_ki_l}=\frac{1}{V}\int\!\! d\vec r\, \psi_{i_k}^2(\vec r) \psi_{i_l}^2(\vec r).
\end{equation} 
They are self-averaging quantities with negligibly small 
fluctuations around their mean $\langle A_{i_ki_l}\rangle=1+2\delta_{i_ki_l}$,
which follows from the independent Gaussian distributions 
for $\psi_i(\vec r)$ \cite{Pri93}. 
Because the correlations between $A_{i_ki_l}$'s and $\gamma_{i_s}$'s
are also negligibly small, we may substitute $A_{i_ki_l}=1+2\delta_{i_ki_l}$
in Eq.~(\ref{lin}). 
Without loss of generality we can assume that 
$\gamma_{i_1}/g(\omega_{i_1})\leq 
\gamma_{i_2}/g(\omega_{i_2})\leq \ldots\leq \gamma_{i_N}/g(\omega_{i_N})$.
 Inverting 
the matrix $A_{i_ki_l}$ we find
\begin{equation}
\frac{g(\omega_{i_k})n_{i_k}}{{\varepsilon} R_{p_0}}=
\frac{1}{N+2}
-\frac{\gamma_{i_k}}{2g(\omega_{i_k})}
+\frac{1}{2(N+2)}\sum_{l=1}^N \frac{\gamma_{i_l}}{g(\omega_{i_l})}.
\label{sol}\end{equation} 
The number of excited modes $N$ is restricted by the requirement 
that all $n_{i_k}$'s should be positive. A necessary and sufficient condition is
\begin{equation}
(2+N)\frac{\gamma_{i_N}}{g(\omega_{i_N})}
-\sum_{l=1}^N\frac{\gamma_{i_l}}{g(\omega_{i_l})}<2.
\label{shcon}\end{equation} 

Eq.~(\ref{shcon}) can be used to determine the 
probability distribution
of the number of excited modes, using the Porter-Thomas distribution 
(\ref{chi2})  for the statistics of decay rates $\gamma_i$. 
In the region of parameters where $\langle N\rangle\gg1$ this mean 
value can be found analytically from the continuous approximation of 
the condition~(\ref{shcon}),
\begin{equation}
\left(2+\int_0^{\alpha_{max}}\!\!\!\!d\alpha\,\sigma(\alpha)\right)\alpha_{max}-
\int_0^{\alpha_{max}}\!\!\!\!d\alpha\,\alpha\sigma(\alpha)=2,
\label{almax}\end{equation}
with $\langle N\rangle = \int_0^{\alpha_{max}}d\alpha\,\sigma(\alpha)$.
The density $\sigma(\alpha)$ of the variables $\alpha_i=\gamma_i/g(\omega_i)$
is given by
\begin{eqnarray}
&&\sigma(\alpha)=\int_0^{\omega_{max}}\!\!\!\!d\omega\,\rho(\omega)
 P_1[\alpha g(\omega)/\bar\gamma(\omega)]g(\omega)/\bar\gamma(\omega)\cr
&&\hphantom{xxxxxxxxxxxx}=
\case{1}{2}\tilde C_1(\Omega/\delta\omega_0)\alpha^{-1/2}.
\end{eqnarray}
It follows from Eq.~(\ref{almax}) that 
$\langle N\rangle=\tilde C_1 (\Omega/\delta\omega_0)z$,
where $z$ satisfies a cubic equation
\begin{equation} 
z^2+\case{1}{3}\tilde C_1 (\Omega/\delta\omega_0)z^3={\varepsilon}.
\label{Nexa}\end{equation} 
To leading order in $1/\langle N\rangle$ the term $z^2$ can be neglected, 
which yields a simple answer
\begin{equation}
\langle N\rangle = 3^{1/3}(\tilde C_1 \Omega/\delta\omega_0)^{2/3}{\varepsilon}^{1/3}=
3^{1/3} \langle N_{nc}\rangle^{2/3}.
\end{equation}
The general form of this result for any $\nu$ can be derived in a similar
way, leading to $\langle N_{nc}\rangle = 
\tilde C_\nu(\Omega/\delta\omega_0){\varepsilon}^{\nu/2}$, 
$\langle N\rangle=(\nu+2)^{\nu/(\nu+2)}\langle N_{nc}\rangle^{2/(\nu+2)}$. 
These results are independent of $\gamma_*$ as long as $\gamma_*\ll\gamma_0$.

\vglue 10truemm
\begin{figure}
\hspace*{15truemm}
\epsfxsize=0.65\hsize
\epsfysize=0.11\vsize
\epsffile{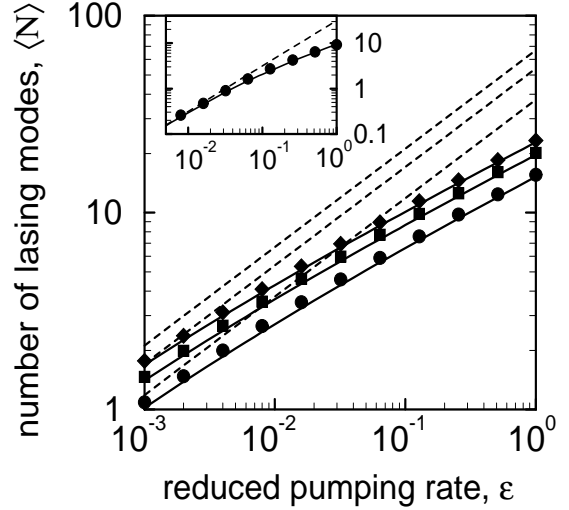}
\vglue 15truemm
\caption{Average number of excited modes $\langle N\rangle$ 
versus dimensionless pumping rate~${\varepsilon}$
(same parameters as in Fig.~1).
The solid lines are the analytical result~(\ref{Nexa}),
the data points are a Monte Carlo average.
The main plot corresponds to $\nu=1$, $M=10^2$ (circles), $M=10^3$ (squares),
$M=10^4$ (diamonds). 
Dashed lines represent the average number $\langle N_{nc}\rangle$
of non-competing modes. 
The inset shows the case $\nu=2$ for $M\gg1$.
Note a drastic reduction 
in the number of excited modes.}
\label{f1}\end{figure}

To test numerically the analytical results for
$\langle N\rangle$, we did a Monte Carlo average over the Porter-Thomas
distribution. 
For each of 2000 realizations, we ordered the modes in increasing order 
of the ratio loss over gain and found maximal $N$ satisfying Eq.~(\ref{shcon}).
Results for $\langle N({\varepsilon})\rangle$ are in excellent agreement with 
the continuous approximation
down to $\langle N\rangle\sim1$ (Fig.~\ref{f1}).

\section{CONCLUSION}
\label{conclusions}

To summarize, we have considered lasing 
of a chaotic cavity coupled to the outside world via
$\nu$ small holes. 
We assumed that the broadening of the cavity modes 
(due to leakage through the holes and absorption 
by the cavity walls) is less
than their spacing and used a simple criterion 
``modal gain~$\geq$~modal loss'' as the condition for a given
mode to be excited. Natural unit of the  
pumping rate $R_{p_0}$ is defined such that 
``maximal gain~$=$~mean loss.'' Because of strong fluctuations
of modal widths, the probability of lasing can be significantly
large for much weaker pumping rates than $R_{p_0}$. The distribution 
of the lasing threshold  turns out to be wide, with the mean much
less than $R_{p_0}$. We have described the multi-mode operation as a result of spatial hole
burning and found that the average number of excited modes
is proportional to the power $\nu/(\nu+2)$ of the pumping rate.

\acknowledgements
This work was supported by the Nederlandse Organisatie voor
Wetenschappelijk
Onderzoek (NWO) and the Stichting voor Fundamenteel Onderzoek der
 Materie (FOM).

\end{multicols}


\begin{references}
\bibitem{Pri93} Spatial correlations in $\psi_i(\vec r)$ exist on the scale
of the wavelength $\lambda$. These are irrelevant for the calculation
of $A_{ij}$ when $D\gg\lambda$.
\end{references}
\end{document}